\documentclass{iopjournal}

\usepackage{amsmath,amssymb}

\usepackage{ulem} 

\begin{document}

\articletype{
Paper
} 

\title{
The sigma meson ($f_0$) at finite temperature with truncated overlap fermions
}

\author{
Susumu~Date$^1$, 
Yuko~Murakami$^2$, 
Motoo~Sekiguchi$^3$, 
Hiroaki~Wada$^3$ and 
Masayuki~Wakayama$^{4,5,*}$
}

\affil{$^1$D3 Center, The University of Osaka, 567-0047, Ibaraki, Osaka, Japan}

\affil{$^2$Information Media Center, Hiroshima University, 739-8511, Hiroshima, Japan}

\affil{$^3$School of Science and Engineering, Kokushikan University, 154-8515, Tokyo, Japan}

\affil{$^4$Department of Physics, Chiba Institute of Technology, 275-0023, Chiba, Japan}

\affil{$^5$Research Center for Nuclear Physics (RCNP), The University of Osaka, 567-0047, Ibaraki, Osaka, Japan}

\affil{$^*$Author to whom any correspondence should be addressed.}

\email{masayuki.wakayama@chibatech.ac.jp}


\def\br#1{\left( #1 \right)}
\def\bra#1{\left\{ #1 \right\}}
\def\brac#1{\left[ #1 \right]}
\def\non{\nonumber}
\def\ds{\displaystyle}

\begin{abstract}
We study the temperature dependence of meson screening masses in two-flavour lattice QCD using dynamical truncated overlap fermions, 
a type of lattice chiral fermions. 
The screening masses for the $\pi$, $\rho$, $a_1$, $a_0$, and the sigma $(f_0)$ mesons are extracted by computing spatial correlation functions.  
This exploratory study uses meson screening masses and the connected/disconnected decomposition of the scalar correlator as diagnostic observables to examine how the disconnected contribution changes across the pseudocritical region.
Above the pseudocritical temperature $T_{\rm pc}$, the $\pi$ and $f_0$ screening masses become degenerate, consistent with chiral restoration. 
The $(\rho,a_1)$ pair also shows the expected degeneracy. 
Decomposition of the $f_0$ propagator reveals that the connected contribution dominates above $T_{\rm pc}$, while the disconnected part becomes significant below $T_{\rm pc}$, explaining the reduced statistical clarity observed at low $T$. 
These results demonstrate that dynamical truncated overlap fermion simulations can capture the qualitative thermal behaviour of the scalar sector. 

\end{abstract}

\section{Introduction}
Determining the theoretical phase diagram for the strong interaction helps clarify the non-trivial vacuum structure, spontaneous symmetry breaking, and confinement mechanisms of gauge theories in terms of control parameters such as temperature and density. 
By deriving the phase diagram from first-principles calculations and clarifying which phases occur and under what conditions, we can assess the validity of effective models such as the Nambu--Jona--Lasinio model~\cite{Nambu:1961tp,Nambu:1961fr,Kunihiro:1991qu,Hatsuda:1994pi} and the hadron resonance gas model~\cite{Hagedorn:1965st,Dashen:1969ep}. 
Furthermore, this approach can locate heavy-ion experiments such as RHIC, LHC, FAIR, and J-PARC-HI within the temperature--density plane. 
Moreover, whether the QCD phase transition in the early universe was first-order or a crossover may lead to different outcomes for baryon-number generation and primordial gravitational-wave signals.

In lattice QCD, in order to investigate meson behaviour during phase transitions at finite temperature, the spatial correlation function is calculated and the screening mass of the meson is extracted from it. 
The meson screening masses reflect changes in meson properties with increasing temperature, such as 
reduced binding or dissociation, and therefore, they are crucial for understanding QCD phase transitions. 
In particular, we are interested in chiral symmetry restoration to verify the validity of the effective model. 
The restoration of chiral symmetry manifests as a degeneracy in the hadron spectral function in the chiral partner channel. 
Under chiral $SU(2)_{\rm L} \times SU(2)_{\rm R}$ symmetry, the chiral partner of the pion is the sigma meson (isoscalar-scalar meson), and the partner of the isovector-vector meson $\rho$ is the isovector-axial vector meson $a_1$. 
In $U(1)_{\rm A}$ symmetry, the pion and the $a_0$ meson are partners. 
Conversely, the degree of degeneracy in the spectral functions of partners can be utilised as a measure of symmetry restoration. 

Calculating the isoscalar-scalar channel (the sigma meson, $f_{0}$) in lattice QCD simulations requires substantial computational resources due to the inclusion of disconnected diagrams, unlike other channels, which only involve connected diagrams. 
This scalar channel is crucial for understanding the origin of meson masses, but the computational cost means that calculations at the physical point are still some way off. 
Here, we summarise the research on sigma mesons in the vacuum and at finite temperature using lattice QCD. 
Several research groups~\cite{Kunihiro:2003yj,Hart:2006ps} have determined the pole masses of sigma mesons. 
Furthermore, as the masses derived from scattering amplitudes are also important for unstable particles with broad decay widths, such as sigma mesons, results have been reported~\cite{Alford:2000mm,Prelovsek:2010kg,Fu:2013ffa,Briceno:2016mjc,Briceno:2017qmb}. 
At finite temperature, several fermion formulations have been used to investigate the QCD transition and the restoration of chiral and axial symmetries for several temperature points and quark masses~\cite{HotQCD:2019xnw,Kotov:2021rah,Aoki:2024uvl,Aoki:2025mue}. 
However, due to significant statistical noise arising from disconnected diagrams, research in this field has not been as systematic as that concerning $\pi$ or $a_0$. 

This study employs the truncated overlap fermion (TOF) action, which is a lattice chiral fermion action, to perform dynamical simulations. 
This represents the first results for $f_{0}$ using a lattice chiral-fermion approach. 
The primary objective of this study is to obtain a signal for $f_{0}$ using the dynamical TOF action, even with a small lattice size. 
This requires identifying the simulation conditions for which the signal can be detected, such as lattice size, lattice spacing, quark mass, and the number of configurations. 
We anticipate that identifying this parameter region will be a milestone for advancing future lattice QCD simulations of $f_{0}$ towards the physical point. 
Consequently, our simulation results are qualitative in nature, subject to certain limitations. 
Using the meson screening mass as a diagnostic measure, and focusing in particular on the temperature dependence of the contribution from disconnected diagrams in the scalar channel, we arrive at the following conclusions. 
These simulations reveal that the screening mass of $f_{0}$ is degenerate with the pion screening mass above the pseudocritical temperature $T_{\rm pc}$, while below $T_{\rm pc}$, this degeneracy lifts and the screening mass becomes comparable to that of the $\rho$ meson. 
Furthermore, it was found that the signal from the connected and disconnected diagrams constituting $f_{0}$ is dominated by the connected diagram above $T_{\rm pc}$, while the contribution from the disconnected diagram becomes significant below $T_{\rm pc}$.

\section{Truncated overlap fermion action}
The truncated overlap fermion (TOF) formulation can be regarded as a finite-$N_5$ realisation of the domain-wall/overlap construction, where $N_5$ is the extent of the fifth direction~\cite{Kaplan:1992bt,Furman:1994ky}. 
We denote the five-dimensional fermion field on the $i$-th slice ($i=1,\cdots,N_5$) by $\psi_i(x)$ with Dirac conjugate $\bar\psi_i(x)$. 
The TOF action is defined as 
\begin{eqnarray}
 S_{\rm TOF} &=& \bar{\psi} D_{\rm TOF} \psi   \, \\
 &=&  \bar{\psi}_1 \brac{ \br{D^{\parallel}-1}\psi_1 + \br{D^{\parallel}+1}P_{\rm R} \psi_2 - m_{\rm f} \br{D^{\parallel}+1} P_{\rm L} \psi_{N_5}  }  \non \\
 &+&  \sum_{i=2}^{N_5-1} \bar{\psi}_i \brac{ \br{D^{\parallel}-1}\psi_i + \br{D^{\parallel}+1}P_{\rm R} \psi_{i+1} + \br{D^{\parallel}+1} P_{\rm L} \psi_{i-1}  }  \non \\
 &+&  \bar{\psi}_{N_5} \brac{ \br{D^{\parallel}-1}\psi_{N_5} - m_{\rm f} \br{D^{\parallel}+1}P_{\rm R} \psi_{1} + \br{D^{\parallel}+1} P_{\rm L} \psi_{N_5-1}  }  \, ,
\end{eqnarray}
where $P_{\rm R/L}=(1\pm\gamma_5)/2$ are the chiral projection operators and $m_{\rm f}$ is the bare quark mass. 
The kernel $D^{\parallel}$ is taken to be the Wilson--Dirac operator with a mass parameter (domain-wall height) $M_5$,
\begin{eqnarray}
 D^{\parallel}(x,y) &=& \br{4-M_5} \delta_{x,y} -\frac{1}{2}\sum_{\mu=\pm 1}^{\pm 4} \br{1-\gamma_\mu} U_{\mu}(x)\delta_{x+\hat\mu,y}  \, . 
\end{eqnarray}

Low-energy observables are described by an effective four-dimensional theory of the light boundary mode (the Furman--Shamir field) obtained after integrating out the heavy modes in the fifth direction. 
In practice, the contribution of the heavy bulk modes is subtracted by introducing the Pauli--Villars (PV) field, i.e.\ the same five-dimensional operator with $m_{\rm f}=1$,
\begin{eqnarray}
D_{\rm PV} = D_{\rm TOF}(m_{\rm f}=1) \, . 
\end{eqnarray}
The effective four-dimensional Dirac operator $D$ is then obtained from the PV-subtracted five-dimensional propagator by projecting onto the boundary/light-field combination. 
Concretely, we define
\begin{eqnarray}
 D &=& \br{P^{\dag} D^{-1}_{\rm PV}D_{\rm TOF} P }_{1,1}   \, , 
 \label{tof_def}
\end{eqnarray}
where the projector $P$ specifies how the four-dimensional light field is embedded into the five-dimensional field,
\begin{eqnarray}
 P_{x_5y_5} &=& P_{\rm L}\delta_{x_5,y_5} + P_{\rm R}\delta_{x_5+1,y_5} + P_{\rm R}\delta_{x_5,N_5} \delta_{y_5,1}   \, . 
\end{eqnarray}
In equation~(\ref{tof_def}), $(\cdots)_{1,1}$ denotes the boundary-to-boundary block in the fifth-direction indices. 
In what follows, we use $D$ in equation~(\ref{tof_def}) as the effective four-dimensional Dirac operator, and $D^{-1}(x,y)$ denotes its quark propagator. 
In the limit $N_5\to\infty$, $D$ approaches the overlap Dirac operator, 
and the associated chiral symmetry becomes the Ginsparg--Wilson symmetry; 
see, e.g., references~\cite{Narayanan:1993zzh,Neuberger:1997fp,KikukawaNoguchi1999,Kikukawa:1999dk,Borici:1999zw}. 

Compared with exact overlap fermions, the finite-$N_5$ TOF formulation avoids the direct treatment of the matrix sign function and provides a computationally tractable five-dimensional representation. 
The trade-off is that chiral symmetry is only approximate at finite $N_5$. 
Compared with other domain-wall formulations, TOF represents a different finite-$N_5$ approximation to the overlap operator, and no universal computational or chiral-symmetry advantage should be assumed without a direct benchmark. 
However, as TOF possesses lattice chiral symmetry, just like DWF, it exhibits better properties with regard to chiral symmetry than Wilson fermions.

\section{Lattice set up}
We used the Wilson gauge action and the TOF action in the calculation at zero and finite temperatures. 
The meson masses at zero temperature were calculated with an $8^3\times 16$ lattice, with $N_5=12$, $M_5 = 1.65$, and $\beta = 5.70$. 
Gauge configurations were generated using the hybrid Monte Carlo (HMC) method. 
After gauge configurations from the first 200 iterations were discarded as thermalisation, configurations were saved every 5 iterations thereafter. 
We stored 345, 350, and 300 configurations for $m_{\rm f} a = 0.10$, 0.09, and 0.08, respectively.

The screening meson masses at finite temperatures were calculated with 
an $8^3\times 4$ lattice, $N_5=12$, and $M_5 = 1.65$. 
A temperature $T$ is defined as $T=1/(N_t a)$ with a temporal lattice size $N_t=4$ and a lattice spacing $a$. 
We varied $a$ by changing $\beta=5.25$, 5.35, 5.45, 5.55, 5.65, 5.70, and 5.75 in order to conduct investigations at various temperatures. 
Our calculations at finite temperatures were performed with $m_{\rm f} a=0.10$, 0.09, and 0.08 for each $\beta$. 
We prepared 80 configurations after 200 iterations as thermalisation for each ($\beta, m_{\rm f} a$) set. 
To determine the temperature dependence, we need to select the $m_{\rm f} a$ corresponding to $\beta$ while maintaining the line of constant physics, which is determined by fixing the $\pi$ and $\rho$ meson mass ratio at zero temperature. 
Lines of constant physics have been investigated, for example, in the context of the clover-improved Wilson fermion action, 
in references~\cite{CP-PACS:2001hxw,Maezawa:2007fc}. 
However, due to the computational cost, the line of constant physics in TOF action has not yet been investigated. 
In this work, we obtain the $\beta$ dependences for several $m_{\rm f} a$ as a rough temperature dependence, 
leaving the determination of the line of constant physics in the TOF action for future work. 
Instead, we provide a representative physical constant at $\beta=5.70$ from simulations at zero temperature. 

The finite-volume effects and other systematic uncertainties associated with the present lattice setup are discussed in section~\ref{sys_err}. 

A TOF code for the simulations was developed based on the Lattice QCD Tool Kit in Fortran 90 (LTKf90)~\cite{Choe:2002pu}. 
The simulations were performed on vector nodes of the SQUID~\cite{Date} at the D3 Center, the University of Osaka, and on the NEC SX-Aurora TSUBASA at Kokushikan University.

\section{Simulation results}

\subsection{Zero temperature}

\begin{figure}[tbp]
\centering
\includegraphics[scale=0.50]{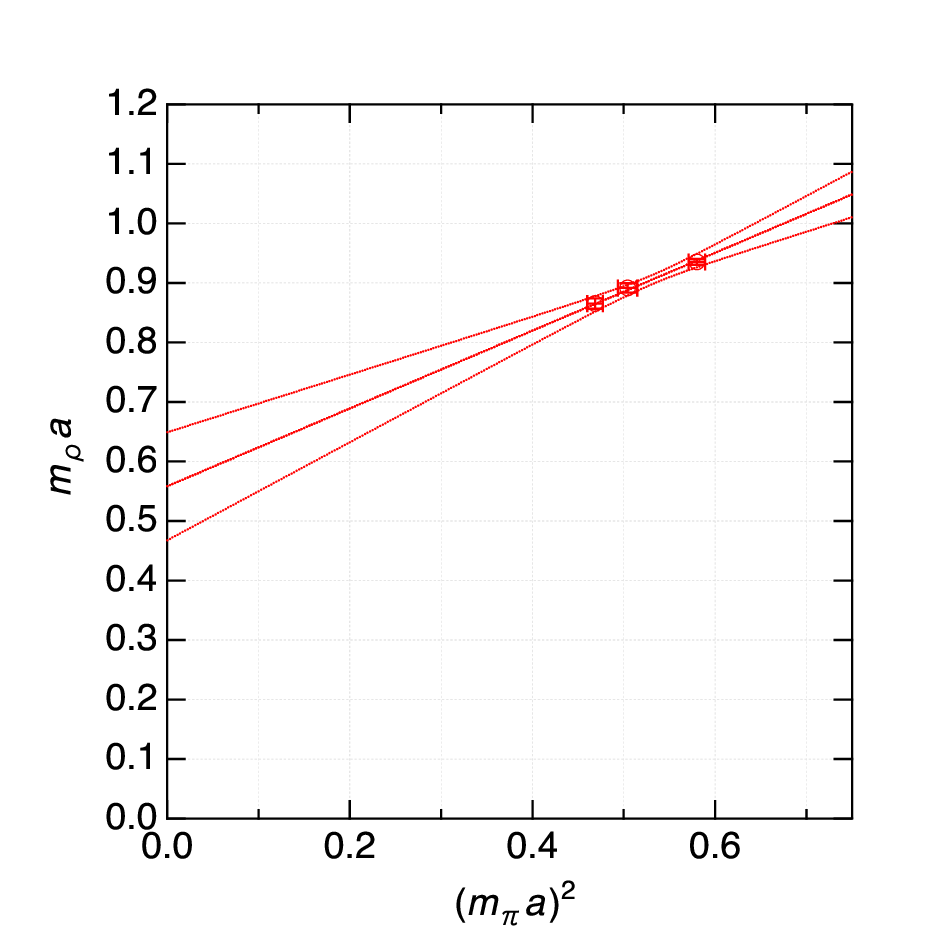}
\caption{\label{fig:mass}
$(m_{\pi}a)^2$ dependence of $\rho$ meson mass. 
}
\end{figure}

Table~\ref{tab:zore_temp} lists the number of configurations, masses of the $\pi$ and $\rho$ mesons, and the mass ratio, $m_{\pi}/m_{\rho}$, for each quark mass for the simulation at zero temperature. 
We estimate the chiral limit from a linear extrapolation of the square of the pion mass, $(m_{\pi}a)^2$, as figure~\ref{fig:mass}. 
From a comparison between the $\rho$ meson mass at the chiral limit, 
$m_{\rho}a=0.558(91)$, 
and the physical mass $m_{\rho}=775.26$~MeV, we obtain the lattice spacing 
$a=0.142(23)$~fm.

\begin{table}[tbp]
\caption{
Number of configurations, masses of the $\pi$ and $\rho$ mesons, and mass ratio, $m_{\pi}/m_{\rho}$ for each quark mass.}
\centering
\begin{tabular}{lc|crc}
\hline
$m_{\rm f} a $ & Confs.  & $m_{\pi} a$ & $m_{\rho} a$ \ \ \  & $m_{\pi}/m_{\rho}$ \\
\hline 
0.10 & 345 & 0.762(6) & 0.935(9) & 0.814(11)  \\
0.09 & 350 & 0.710(7) & 0.891(8) & 0.796(15)  \\
0.08 & 300 & 0.684(6) & 0.865(5) & 0.791(15)  \\
\hline
\end{tabular}
\label{tab:zore_temp}
\end{table}

\begin{figure}[tbp]
\centering
\includegraphics[scale=0.50]{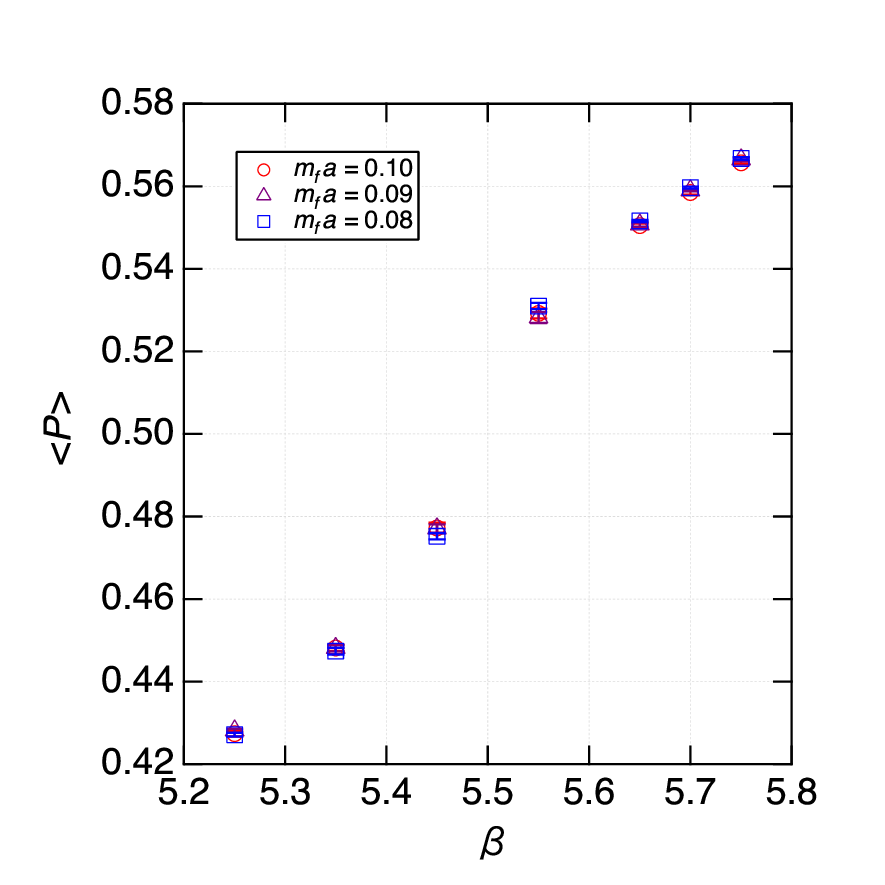}
\hfill
\includegraphics[scale=0.50]{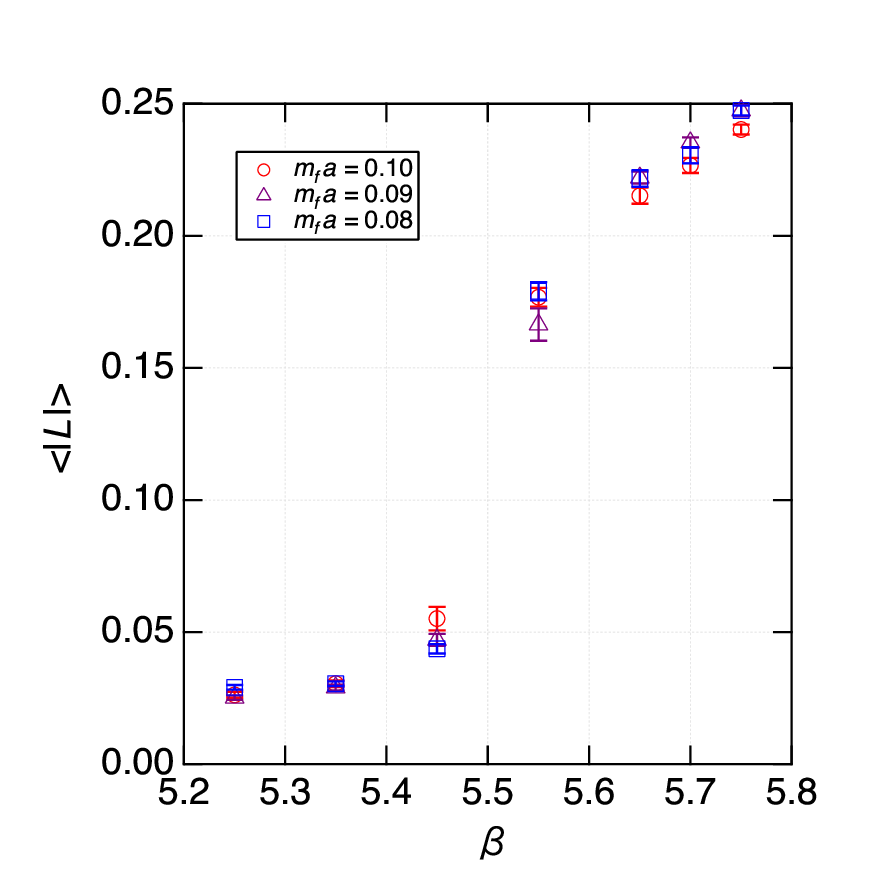}
\caption{\label{fig:gauge}
$\beta$ dependence of expectation value of plaquettes (left panel) and expectation value of absolute value of Polyakov loops (right panel) in lattice units. 
}
\end{figure}

Figure~\ref{fig:gauge} presents the $\beta$ dependences of the expectation values of plaquettes $\left \langle P \right \rangle$ and the absolute value of Polyakov loops $\left \langle |L| \right \rangle$. 
The plaquette $P$ is defined by 
\begin{eqnarray}
 P &=& \frac{1}{12N_v N_c}\sum_{x}\sum_{ \substack{\mu,\nu=1\\ \mu\neq\nu} }^{4}{\rm Re \, }{\rm Tr}_c\brac{U_{\mu}(x) U_{\nu}(x+\hat\mu) U_{\mu}^{\dag}(x+\hat\nu) U_{\nu}^{\dag}(x)} \, ,
\end{eqnarray}
where $U_{\mu}(x)$ is the link variable, $\hat \mu$ is the unit vector along the $\mu$ direction, ${\rm Tr}_c$ is the trace over the colour index, $N_v$ is the number of lattice space--time sites, and $N_c(=3)$ is the number of colour degrees of freedom. 
The Polyakov loop $L$ is defined by 
\begin{eqnarray}
 L &=& \frac{1}{N_sN_c}\sum_{\vec{x}}{\rm Tr}_c  \prod_{t=0}^{N_t-1} U_{4}(t,\vec{x}) \, ,
\end{eqnarray}
where $N_s$ is the number of spatial lattice sites.

The behaviour of the expectation value in figure~\ref{fig:gauge} shows that the simulations above $\beta = 5.65$ lie above the pseudocritical temperature $T_{\rm pc}$ of the chiral phase transition and those below $\beta=5.35$ lie below $T_{\rm pc}$. 
As a representative physical temperature, we consider 
$T=347(56)$~MeV 
at $\beta=5.70$ from the lattice spacing 
$a=0.142(23)$~fm with $T=1/(N_t a)$.

\subsection{Finite temperature}

\begin{figure}[tbp]
\centering
\includegraphics[scale=0.50]{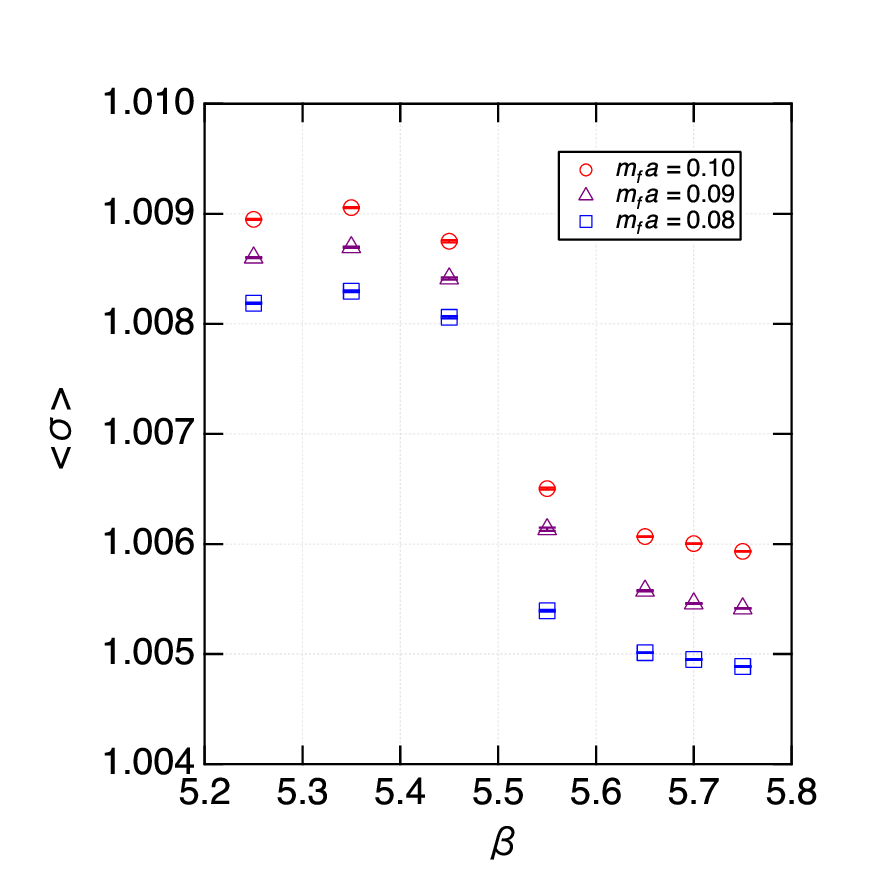}
\caption{\label{fig:sigma}
$\beta$ dependence of $\left \langle \sigma \right \rangle$ 
in lattice units. 
}
\end{figure}

Next, we calculate the spatial propagators. 
In constructing the mesons, we adopt the interpolating operators $M_i(x)$ ($i=\pi, a_0, f_0, \rho, a_1$) with simple point-like sources and sinks 
as given in table~\ref{tab:meson}. 
\begin{table}[tbp]
\caption{
Meson states, corresponding operators $M_i(x)$ ($i=\pi, a_0, f_0, \rho, a_1$), isospin, and $J^{PC}$, 
where $J$, $P$, and $C$ are total angular momentum, parity, and charge, respectively. 
$\tau^a$ ($a=1,2,3$) is the Pauli matrix representing the flavour $SU(2)_{\rm f}$. 
$q$ is the iso-doublet quark field, $q=(u,d)^{T}$. 
}
\centering
\begin{tabular}{ccccc}
\hline
State & & \hspace{-25mm}$M_i(x)$: Operator  & Isospin & $J^{PC}$  \\
\hline 
$\pi$ & $ M_{\pi\ }(x)=$&\hspace{-3.5mm}$\bar{q}(x) \gamma_5 \frac{\tau^a}{2} q(x)$ & 1 & $0^{-+}$   \\
$a_0$ & $ M_{a_0}(x)=$&\hspace{-3.5mm}$\bar{q}(x) \frac{\tau^a}{2} q(x)$  & 1& $0^{++}$   \\
$f_0$ & $ M_{f_0}(x)=$&\hspace{-3.5mm}$\bar{q}(x)  q(x)$  & 0 & $0^{++}$   \\
$\rho$ & $M_{\rho\ }(x)=$&\hspace{-3.5mm}$\bar{q}(x) \gamma_i \frac{\tau^a}{2} q(x)$  & 1 & $1^{--}$  \\
$a_1$ & $M_{a_1}(x)=$&\hspace{-3.5mm}$\bar{q}(x) \gamma_i \gamma_5 \frac{\tau^a}{2} q(x)$  & 1 & $1^{++}$  \\
\hline
\end{tabular}
\label{tab:meson}
\end{table}
The spatial meson propagators are defined as  
 \begin{eqnarray}
 G_{i}(x) &=& \left \langle M_{i}(x)M_{i}^{\dag}(0) \right \rangle   \, .
 \label{propa_def}
\end{eqnarray}
For $i=\pi, a_0, \rho,$ and $a_1$, equation~(\ref{propa_def}) is written as 
 \begin{eqnarray}
 G_{i}(x) &=&  - \left \langle {\rm Tr}\brac{\Gamma_i D^{-1}(x,0) \Gamma_i D^{-1}(0,x)} \right \rangle   \, ,
 \label{conn}
\end{eqnarray}
where 
$D^{-1}(x,y)$ denotes the quark propagator from $x$ to $y$, ${\rm Tr}$ is the trace taken over the colour and Dirac indices, 
and $\Gamma_i = \gamma_5, 1, \gamma_i,$ and $\gamma_i\gamma_5$, respectively. 
In contrast, for $i=f_0$,  equation~(\ref{propa_def}) is written as 
 \begin{eqnarray}
 G_{f_0}(x) &=&  - \left \langle {\rm Tr}\brac{D^{-1}(x,0) D^{-1}(0,x)} \right \rangle \non \\
 &&      + 2 \left \langle {\rm Tr} \brac{ D^{-1}(x,x)}  {\rm Tr} \brac{ D^{-1}(0,0)} \right \rangle  \, .
 \label{disc}
\end{eqnarray}
Equation~(\ref{conn}) and the first term of equation~(\ref{disc}) correspond to the so-called connected diagrams, 
and the second term of equation~(\ref{disc}) corresponds to the so-called disconnected diagrams. 
The disconnected diagram vanishes except for in the $f_0$ meson calculation. 
It is not easy to evaluate the disconnected diagram, since we have to calculate ${\rm Tr} \brac{D^{-1}(x,x)}$, that is $\sigma(x)$, for all lattice sites. 
The $Z_2$-noise method~\cite{McNeile:2000xx,TXL:2000mhy} with or without the truncated eigenmode approach~\cite{Neff:2001zr} is usually used to calculate the disconnected diagram. 
However, we evaluate $\sigma(x)$ by directly computing the inverse of the Dirac matrix at all lattice sites to avoid contamination from fluctuations in the noise method; see section~\ref{z2_noise} later. 
To improve accuracy, we use space-time symmetries, such as space-time translations and cubic rotations, to exploit the gauge configurations effectively. 
The isoscalar-scalar channel at zero momentum has a vacuum contribution; we account for this by evaluating vacuum-subtracted correlations. 
Namely, we subtract 
$\left \langle \sigma \right \rangle$ 
of the disconnected diagram to extract its signal~\cite{Kunihiro:2003yj,Hart:2006ps,Wakayama:2014gpa}: 
 \begin{eqnarray}
 \left \langle \sigma(x)\sigma(0) \right \rangle 
 &=& \left \langle \br{\sigma(x)  - \left \langle \sigma \right \rangle } \br{ \sigma(0) - \left \langle \sigma \right \rangle } \right \rangle  \, , 
\end{eqnarray}
where the lattice data of $\left \langle \sigma \right \rangle$ in the simulations are shown in figure~\ref{fig:sigma}. 

\begin{figure}[tbp]
\centering
\includegraphics[scale=0.50]{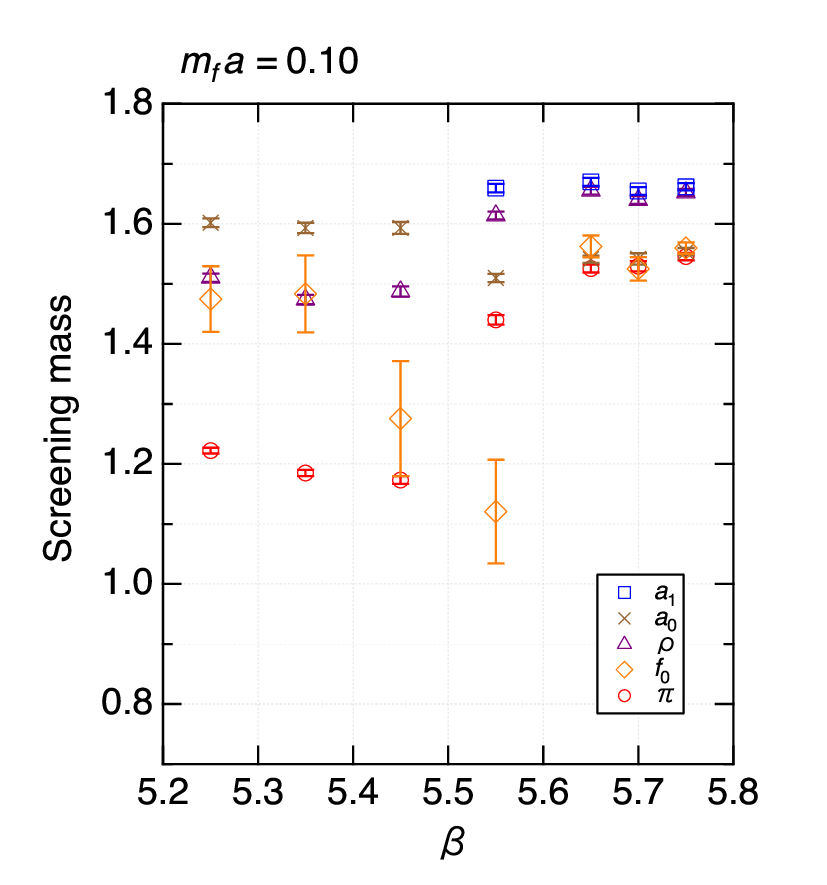}
\hfill
\includegraphics[scale=0.50]{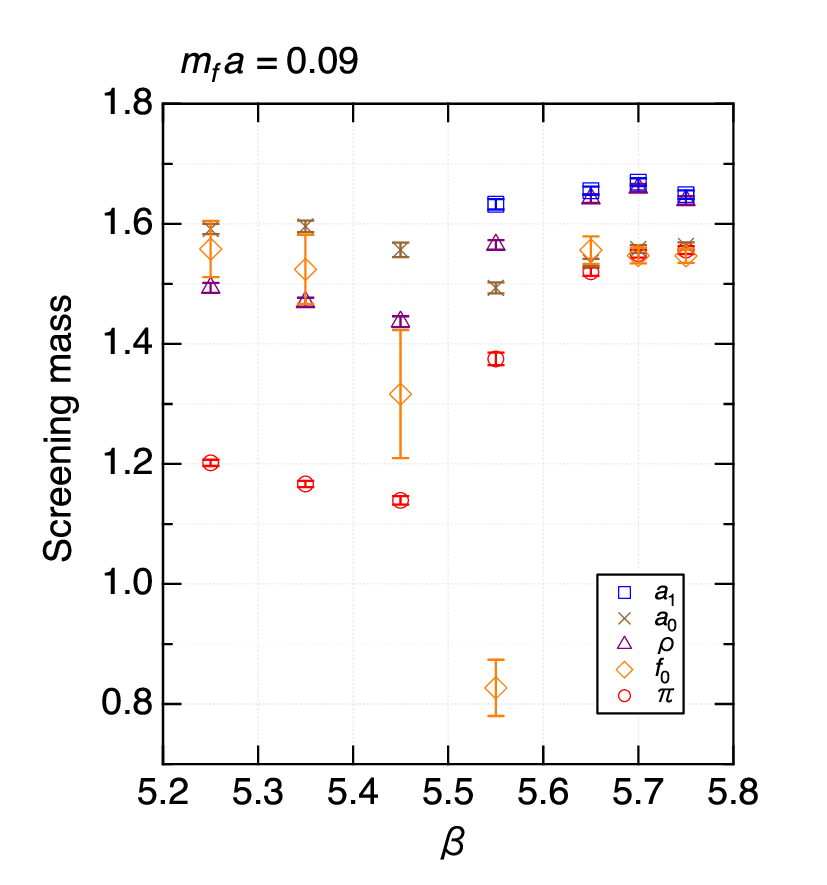}
\hfill
\includegraphics[scale=0.50]{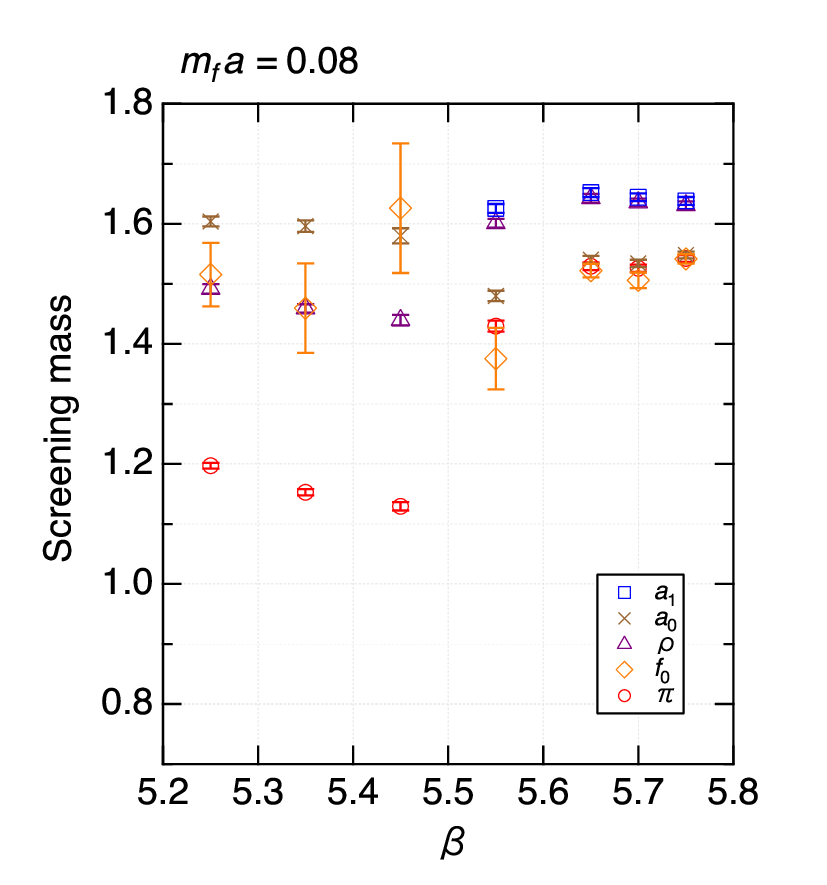}
\caption{\label{fig:eff_mass}
$\beta$ dependence of screening masses of the mesons at $m_{\rm f} a=0.10$, 0.09, and 0.08 in lattice units. 
}
\end{figure}

We obtain the screening masses for the $\pi$, $\rho$, $a_1$, $a_0$, and $f_0$ mesons at finite temperature from the spatial propagators. 
Figure~\ref{fig:eff_mass} shows the $\beta$ dependences of the screening masses. 
We find that under $T_{\rm pc}$, such as for $\beta=5.25$ and 5.35, the screening masses of the $f_0$ meson are larger than those of the $\pi$ meson and are comparable to those of the $\rho$ meson. 
It is noteworthy that the screening masses of the $f_0$ meson are smaller than those of the $a_0$ meson.
The observed ordering of the $f_0$ and $a_0$ screening masses is qualitatively consistent with the expected low-temperature hierarchy.
For the $a_1$ meson, no signal was obtained below $T_{\rm pc}$ at this lattice setup due to the $a_1$ meson being too heavy.

We find that above $T_{\rm pc}$, such as for $\beta=5.65,$ 5.70, and 5.75, 
the screening masses of the $\pi$ and $f_0$ mesons become degenerate. 
The lattice result supports the scenario predicted by the effective theory, in which the pion, as the Nambu-Goldstone boson, and the sigma meson, as the chiral partner of the pion, become degenerate in the region where the spontaneous breaking of chiral symmetry recovers~\cite{Hatsuda:1984jm,Hatsuda:1985eb,Hatsuda:1985ey,Hatsuda:1986gu}. 
We also find that the screening masses of the $f_0$ and $a_0$ mesons, and the $\rho$ and $a_1$ mesons, become degenerate above $T_{\rm pc}$. 
Regarding the screening masses of the $f_0$ meson, the system offers good accuracy above $T_{\rm pc}$, but not so good near and below $T_{\rm pc}$. 
Note that each data point was calculated using 80 gauge configurations.

\begin{figure}[tbp]
\centering
\includegraphics[scale=0.50]{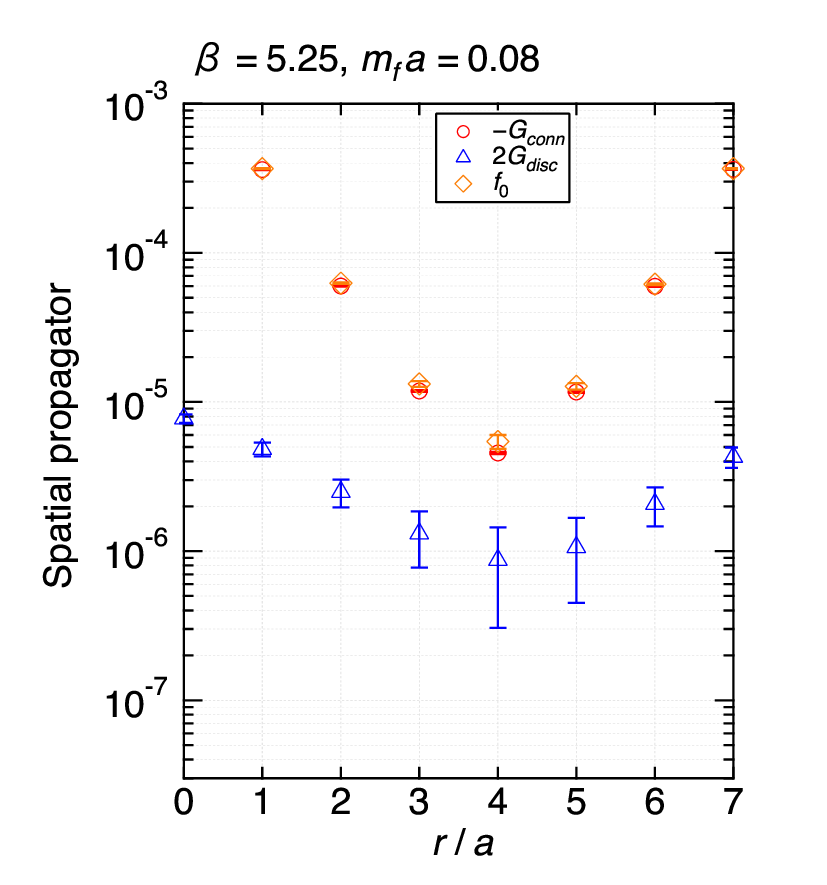}
\hfill
\includegraphics[scale=0.50]{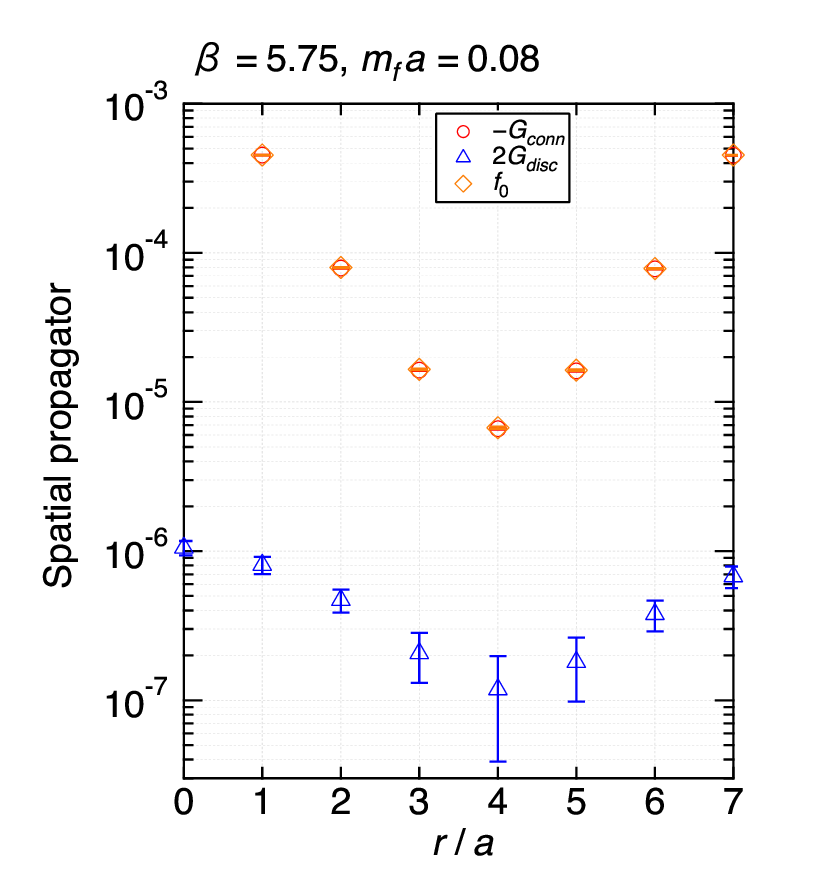}
\caption{\label{fig:conn_disc}
Contribution of connected and disconnected elements to spatial propagators of $f_0$ meson at $\beta =5.25$ and 5.75 in lattice units. 
}
\end{figure}

To further illustrate the degeneration of the $f_0$ and $a_0$ meson, we show the spatial propagators in figure~\ref{fig:conn_disc}. 
The figure shows the spatial propagator of the $f_0$ meson together with its components, the connected diagram $G_{\rm conn}$ and the disconnected diagram $G_{\rm disc}$. 
$G_{\rm conn}$ corresponds to the spatial propagator of the $a_0$ meson. 
Above $T_{\rm pc}$ (for example, at $\beta=5.75$), the spatial propagator of the $f_0$ meson becomes degenerate with that of the connected diagram, namely with the spatial propagator of the $a_0$ meson. 
In this case, the component of the connected diagram is more than about 30 times larger than that of the disconnected diagram. 
Therefore, the contribution of the disconnected diagram to the spatial propagator of the $f_0$ meson is small, and the accuracy of the spatial propagator becomes good. 

On the other hand, below $T_{\rm pc}$ (for example, at $\beta=5.25$), the component of the disconnected diagram can approach about one third of that of the connected diagram, and the accuracy of the spatial propagator becomes poor. 
This shows that at low temperatures, the disconnected diagram plays an important role in the spatial propagator of the $f_0$ meson, making it difficult to obtain a clear signal for the $f_0$ meson. 
This result is consistent with the suggestion obtained from a lattice calculation at zero temperature~\cite{Kunihiro:2003yj,Hart:2006ps}.

\subsection{Comparison of the direct all-site calculation and the $Z_2$-noise method}\label{z2_noise}
\begin{figure}
\centering
\includegraphics[scale=0.60]{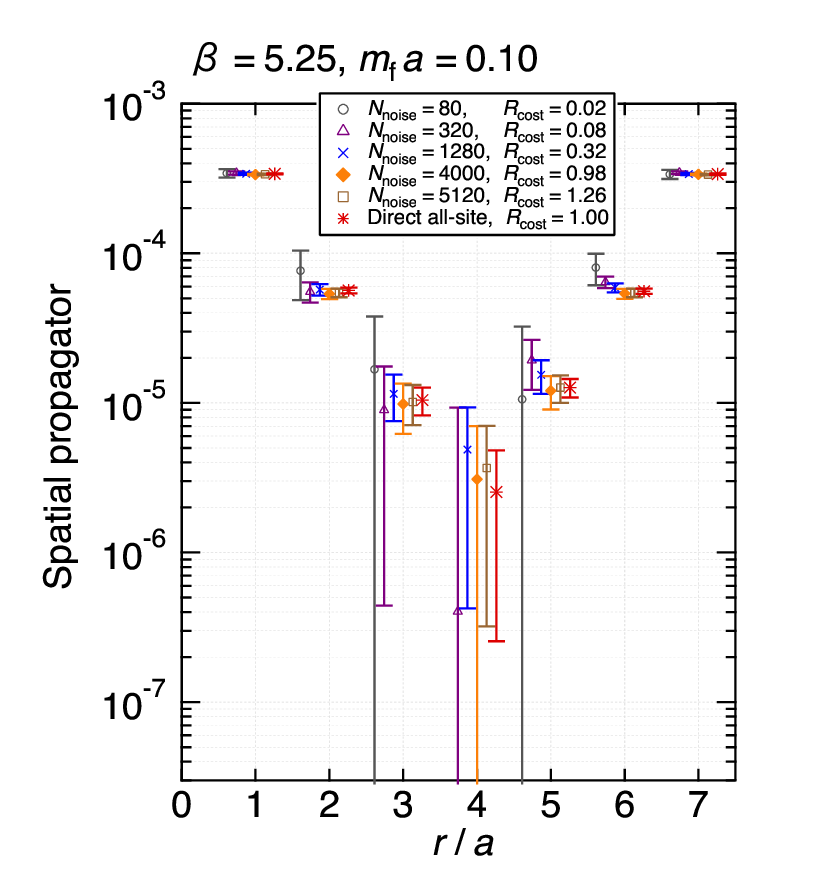}
\caption{\label{fig:noise}
Comparison of the $f_0$ spatial propagator obtained using the $Z_2$-noise method and the direct all-site calculation at $\beta=5.25$ and $m_{\rm f} a=0.10$. 
Here, $N_{\rm noise}$ denotes the number of noise vectors, and $R_{\rm cost}=C_{\rm noise}/C_{\rm direct}$ is the computational cost relative to that of the direct calculation. 
The data points corresponding to different methods are slightly displaced horizontally for visibility. 
}
\end{figure}

To assess the computational efficiency of the direct all-site calculation, we compare it with the $Z_2$-noise method at $\beta=5.25$ and $m_{\rm f} a=0.10$. 
This parameter point lies below the pseudocritical region, where the disconnected contribution to the $f_0$ spatial propagator is significant, and therefore provides a demanding test of the stochastic estimator. We performed the stochastic calculation using $N_{\rm noise}=80$, 320, 1280, 4000, and 5120 noise vectors. 
We define the relative computational cost as 
\begin{eqnarray}
 R_{\rm cost}=\frac{C_{\rm noise}}{C_{\rm direct}} \, , 
 \label{cost_ratio}
\end{eqnarray}
where $C_{\rm noise}$ and $C_{\rm direct}$ denote the measured wall-clock times required to evaluate the $f_0$ spatial propagator using the stochastic and direct methods, respectively, for the same set of 80 gauge configurations. 
The timing measurements were performed using the same computational resources and parallelisation setup, and $R_{\rm cost}=1$ corresponds to the cost of the direct calculation. 
Figure~\ref{fig:noise} compares the resulting $f_0$ spatial propagators. 
At short distances, the stochastic estimates are consistent with the direct result even for relatively small $N_{\rm noise}$. 
At intermediate and large distances, however, where the propagator becomes small, the stochastic uncertainties are substantial for $R_{\rm cost}\ll1$. 
The uncertainties decrease as $N_{\rm noise}$ is increased, but a precision comparable to that of the direct calculation is obtained only when $N_{\rm noise}=4000$ (corresponding to $R_{\rm cost}=0.98$) or larger. 
Thus, for the present lattice setup and the precision required for the $f_0$ spatial propagator, the stochastic estimator does not provide a substantial computational advantage over the direct all-site calculation. 

We emphasise that this conclusion concerns the $Z_2$-noise estimate of the disconnected contribution to the $f_0$ correlator in the present setup. 
On larger lattices, the signal at longer spatial separations decreases exponentially, whereas the stochastic error decreases only as $1/\sqrt{N_{\rm noise}}$; maintaining a comparable relative precision may therefore require a rapidly increasing number of noise vectors. 
The direct method avoids this additional stochastic error, although its cost increases with the lattice volume and gauge-configuration fluctuations also remain. 
However, this expected behaviour has not been tested in the present study, and the relative efficiency of the two methods for disconnected correlators on larger lattices requires further numerical investigation. 
For propagators containing only connected contractions, by contrast, the stochastic $Z_2$-noise method remains computationally advantageous.

\begin{figure}
\centering
\includegraphics[scale=0.50]{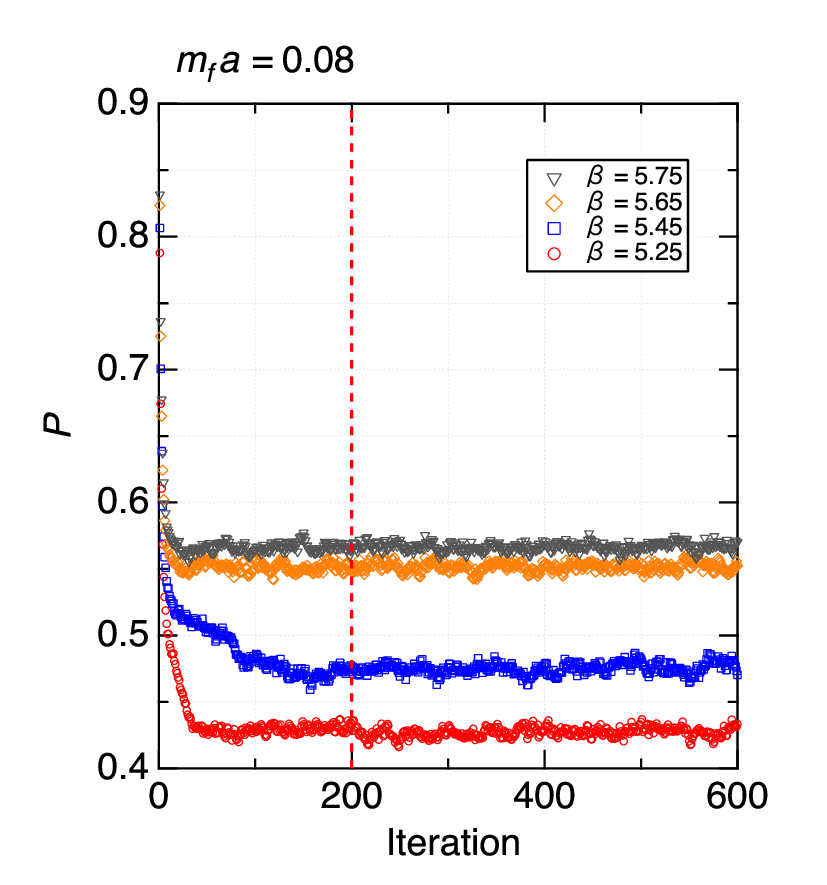}
\hfill
\includegraphics[scale=0.50]{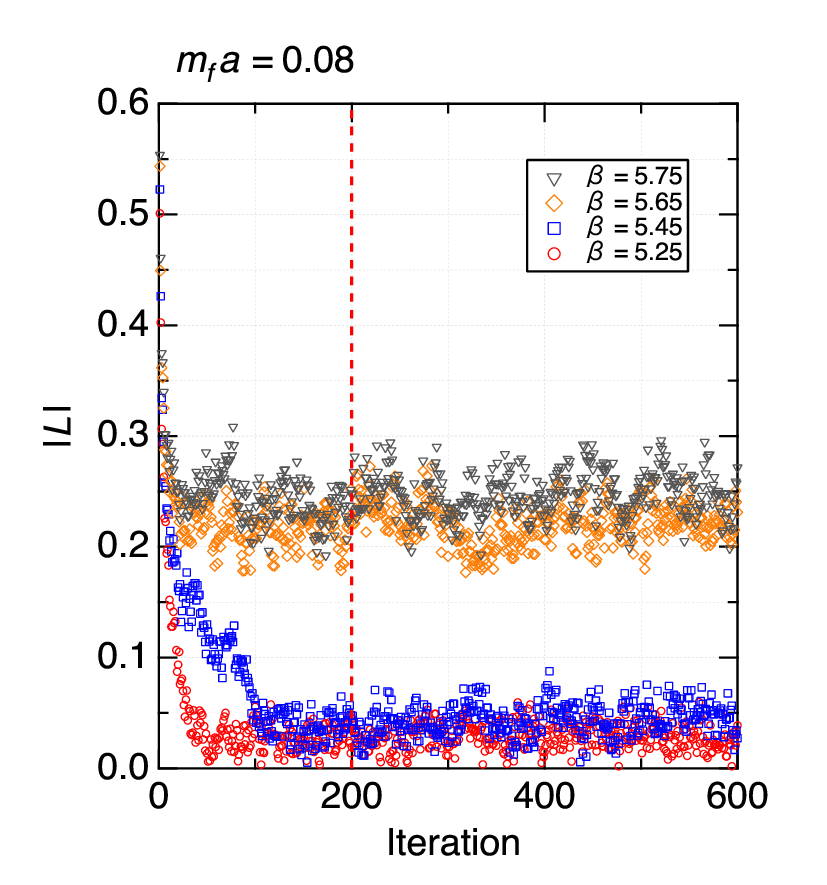}
\caption{\label{fig:history}
HMC histories of the plaquette $P$ (left panel) and the absolute value of the Polyakov loop $|L|$ (right panel) for $m_{\rm f} a = 0.08$ at representative values of $\beta$. 
The vertical dashed line indicates the thermalisation cut at 200 HMC iterations. 
}
\end{figure}

\begin{figure}
\centering
\includegraphics[scale=0.50]{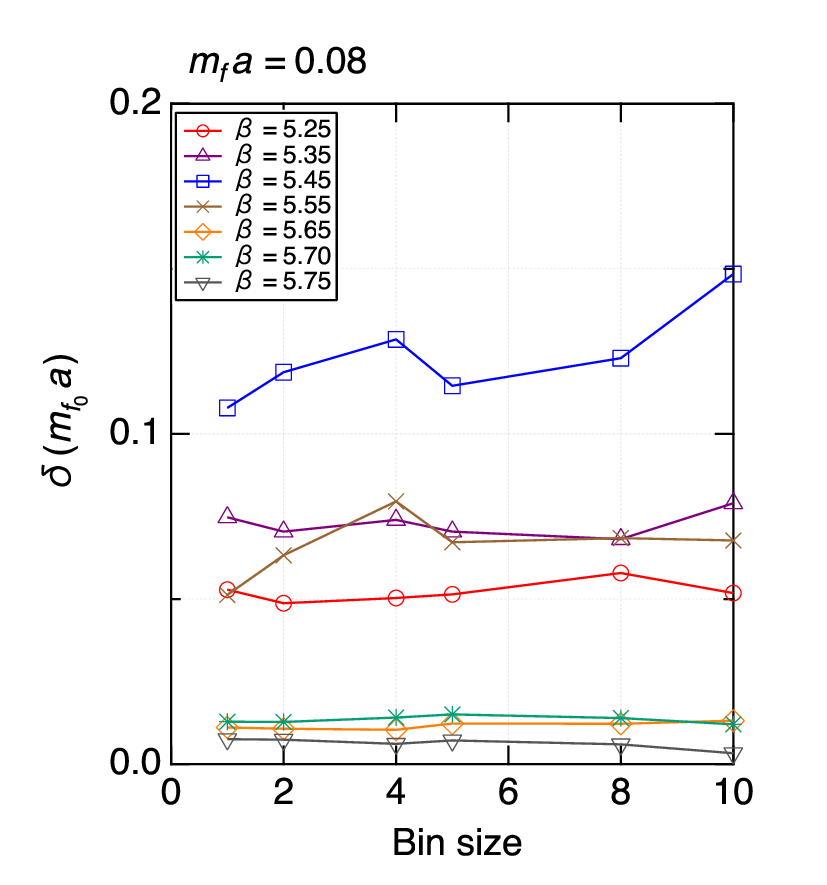}
\caption{\label{fig:bin-size}
Bin-size dependence of the jackknife statistical errors of the $f_0$ screening masses at $m_{\rm f} a = 0.08$. 
The analysis is based on 80 gauge configurations for each value of $\beta$. 
}
\end{figure}

\subsection{Thermalisation and statistical stability}

To examine gauge-ensemble equilibration, we monitored the HMC histories of the plaquette and the absolute value of the Polyakov loop. 
Figure~\ref{fig:history} shows representative histories for the lightest quark mass, $m_{\rm f} a = 0.08$, at several values of $\beta$. 
An initial relaxation is visible, particularly at $\beta=5.45$, which lies near the pseudocritical region in the present lattice setup. 
Both observables, however, reach stationary distributions before the thermalisation cut of 200 HMC iterations, 
and no persistent drift is observed in the subsequent measurement region.

To investigate possible autocorrelation effects on the $f_0$ screening masses, 
we repeated the jackknife error analysis using several bin sizes. 
The $f_0$ channel is statistically more demanding than the flavour-nonsinglet meson channels 
because it contains a disconnected contribution with large configuration-to-configuration fluctuations. 
We therefore use this channel as a conservative test case for the bin-size analysis. 
Figure~\ref{fig:bin-size} shows the bin-size dependence of the statistical errors at $m_{\rm f} a = 0.08$. 
Within the range of moderate bin sizes for which a sufficient number of bins remains, the estimated errors remain broadly stable, 
although a mild increase is visible at $\beta=5.45$. 
This behaviour may be associated with the proximity of this coupling to the pseudocritical region.

Since successive stored configurations are separated by five HMC iterations and the errors obtained with a bin size of one configuration are consistent, within the resolution of the present statistics, we use a bin size of one configuration in the final jackknife analysis.
The relatively weak bin-size dependence observed in the present calculation may partly reflect the small spatial volume and the relatively heavy quark masses, which limit the development of long-range fluctuations. 
Future simulations at larger volumes and lighter quark masses, particularly near the pseudocritical region, will require more careful autocorrelation analyses.

\subsection{Systematic uncertainties}\label{sys_err}
Several systematic uncertainties should be kept in mind when interpreting the present results. 
First, the physical scale has been estimated only at $\beta=5.70$, yielding $a=0.142(23)$ fm and $T=347(56)$ MeV. 
The uncertainty in the lattice spacing propagates directly to the quoted physical temperature and spatial extent. 
Moreover, a line of constant physics has not yet been established for the TOF action. 
Consequently, changing $\beta$ at fixed bare quark mass changes not only the temperature but also the physical quark mass and other dimensionful quantities. 
Therefore, the observed $\beta$ dependence cannot be interpreted as a quantitatively controlled temperature dependence at fixed physics. 

Second, finite-volume effects cannot be quantified because only one spatial volume, $8^3$, has been used. 
At the reference point $\beta=5.70$, the lattice spacing determined from the zero-temperature calculation gives a spatial extent of $L=1.14(18)$~fm. 
Using the corresponding zero-temperature pion masses, we obtain $m_\pi L=6.10(5)$, 5.68(6), and 5.47(5) for $m_{\rm f}a=0.10$, 0.09, and 0.08, respectively. 
Although these conventional indicators suggest that finite-volume effects associated with the pion correlation length may not be exceptionally large at this reference point, they do not ensure small finite-volume effects. 
The limited spatial extent restricts the large-distance region available for extracting screening masses, and the $f_0$ channel may be particularly sensitive to finite-volume effects because of its disconnected contribution and scalar-channel dynamics.
Such effects may also become more important near the pseudocritical region, where longer-range fluctuations can occur. 
Since calculations at different spatial volumes are not available, these effects cannot be estimated quantitatively. 
Furthermore, the physical spatial extent and $m_\pi L$ cannot be evaluated throughout the $\beta$ scan because the scale and zero-temperature hadron masses have not been determined at the other $\beta$. 

Third, all finite-temperature calculations were performed at the single temporal lattice size $N_t=4$. 
Possible discretisation effects on the screening masses therefore cannot be quantified, and a continuum extrapolation cannot be performed. 
Together with the scale-setting and finite-volume uncertainties discussed above, this means that the present results should be interpreted qualitatively.

\section{Summary}
In this study, we investigated the temperature dependence of the $\pi$, $\rho$, $a_1$, $a_0$, and $f_0$ mesons using dynamical truncated overlap fermions (TOF). 
Our primary goal was to identify parameter regions in which a clear signal for the $f_0$ meson can be obtained, even on a small lattice, as a first step toward future simulations approaching the physical point. 

Since the line of constant physics for the TOF action has not yet been established, we investigated the $\beta$ dependence of physical observables for several $m_{\rm f} a$ as a rough temperature dependence. 
As a reference point, we provided a representative physical scale at $\beta=5.70$, where the zero-temperature calculations of the $\pi$ and $\rho$ meson masses yield a lattice spacing of 
$a=0.142(23)$~fm. 
This allows us to relate our finite-temperature simulations to physical units and to estimate that the system at $\beta=5.70$ corresponds to 
$T=347(56)$~MeV. 

In the finite-temperature calculations, the behaviour of observables such as plaquette values and Polyakov loops 
indicates that simulations with $\beta \gtrsim 5.65$ lie above the pseudocritical temperature $T_{\rm pc}$, while those with $\beta \lesssim 5.35$ lie below $T_{\rm pc}$.

The screening masses extracted from the spatial propagators exhibit a characteristic behaviour across $T_{\rm pc}$. 
Below $T_{\rm pc}$, the screening mass of the $f_0$ meson is larger than that of the $\pi$ meson and comparable to that of the $\rho$ meson, while remaining smaller than that of the $a_0$ meson. 
Above $T_{\rm pc}$, the screening masses of the $\pi$ and $f_0$ mesons become degenerate, consistent with the restoration of chiral symmetry in which the pion and sigma meson form a chiral pair. 
We also observe the expected degeneracy of the $(\rho, a_1)$ mesons in the chirally restored regime. 

Above $T_{\rm pc}$, 
the screening masses of the $f_0$ and $a_0$ mesons become degenerate. 
This degeneracy is compatible with a reduction of $U(1)_{\rm A}$ breaking, but screening-mass degeneracy alone does not establish effective $U(1)_{\rm A}$ restoration.
By decomposing the $f_0$ spatial propagator into connected and disconnected parts, we can understand why the behaviour of the screening mass of the $f_0$ meson changes across $T_{\rm pc}$. 
The connected contribution dominates the $f_0$ channel above $T_{\rm pc}$, exceeding the disconnected contribution by more than an order of magnitude; consequently, the $f_0$ meson signal is statistically robust in this regime. 
Below $T_{\rm pc}$, however, the disconnected diagram contributes significantly, approaching the magnitude of the connected component. 
This increases the statistical noise and makes it difficult to obtain a clear signal. 
This observation is consistent with previous zero-temperature findings that emphasise the relevance of the disconnected diagram in the isosinglet scalar channel. 
Within the limitations of the present calculation, this provides qualitative evidence that the disconnected contribution to the scalar channel changes across the pseudocritical region. 
However, as our results do not take into account the effect of the residual mass in TOF, they are strictly qualitative. 
Furthermore, in order to confirm the restoration of the $U(1)_{\rm A}$ symmetry, it is necessary to examine not only the screening mass but also the topological susceptibility and the Dirac eigenvalue spectrum simultaneously; see for review~\cite{Ding:2026gao}.

These results demonstrate that dynamical TOF simulations can reproduce the expected qualitative features of meson screening masses across $T_{\rm pc}$ and can provide signals for the $f_0$ meson under suitable conditions. 
An important task for future work is to determine the line of constant physics for the TOF action. 
In addition, to advance quantitative studies near the physical point, it will be necessary to increase the lattice volume and statistical precision, and to further reduce the quark masses in future simulations.

\ack{
The simulations reported in this paper were performed using
the SQUID supercomputers at the D3 Center of the University of Osaka,
with the support of the Research Center for Nuclear Physics of the University of Osaka, and
using the SX-Aurora TSUBASA system at Kokushikan University. 
This work was
also supported by the Joint Usage/Research Center for Interdisciplinary Large-scale
Information Infrastructures, Japan (project ID jh240034). 
}

\data{
It may be used for future publications. The data that support the findings of this study may be available upon reasonable request from the authors.
}

\end{document}